\documentclass[12pt]{article}
\usepackage[english]{babel}
\usepackage[utf8x]{inputenc}
\usepackage{amsmath}
\usepackage{graphicx}
\usepackage{etoolbox}
\usepackage{changepage}
\usepackage{titlesec}
\usepackage[parfill]{parskip}
\usepackage[margin=1in]{geometry}
\usepackage{times}
\usepackage{float}
\usepackage{ifthen}

\usepackage{cite}

\usepackage{multirow}
\usepackage[dvipsnames]{xcolor}
\usepackage{subcaption}
\usepackage{titlesec}
\usepackage{makecell}

\usepackage{comment}

\titleformat*{\section}{\normalsize\bfseries} 

\titleformat*{\subsection}{\normalsize\it} 

\newenvironment{iquote}
    {\vspace{.4\baselineskip}\itshape\list{}{\leftmargin=0.15in\rightmargin=0.15in}%
    \item\relax}
    {\endlist\vspace{.4\baselineskip}}
    
\newenvironment{tightquote}
    {\vspace{0\baselineskip}\itshape\list{}{\leftmargin=0.15in\rightmargin=0.15in}%
    \item\relax}
    {\endlist\vspace{0\baselineskip}}

\newboolean{redactswitch}
\setboolean{redactswitch}{true} 

\newcommand{\redact}[1]{\ifthenelse{
    \boolean{redactswitch}}{{[Redacted]}}{
    {#1}}}

\newcommand{\fracroot}[2]{\ifthenelse{#1=1}{\frac{1}{\sqrt{#2}}}{\sqrt{\frac{#1}{#2}}}}

\usepackage{makecell}

\newcommand{\who}[1]{\textbf{#1\emph{:}}}

\newboolean{showtimestamp}
\setboolean{showtimestamp}{false}

\newcommand{\timestamp}[1]{
    \ifthenelse{\boolean{showtimestamp}}{\textcolor{red}{Timestamp: #1}}{}}

\usepackage{hyperref}  
\hypersetup{colorlinks=true,urlcolor=blue,citecolor=blue,linkcolor=blue}   
\usepackage{enumitem}        
\setlist{nosep}

\title{\large \textbf{Ethics education in the quantum information science classroom: \\Exploring attitudes, barriers, and opportunities}} 
\author{Josephine Meyer, Noah Finkelstein, and Bethany Wilcox}
\date{\small{Department of Physics, University of Colorado Boulder \\
Submitted to the 2022 ASEE Annual Conference \& Exposition, Minneapolis, MN, USA. Copyright © 2022, American Society for Engineering Education}} 

\makeatletter 
\patchcmd{\@maketitle}{\begin{center}}{\begin{adjustwidth}{0.5in}{0.5in}\begin{center}}{}{}
\patchcmd{\@maketitle}{\end{center}}{\end{center}\end{adjustwidth}}{}{}
\makeatother

\begin{document}
\raggedright
\maketitle

\section*{Abstract}

Quantum information science (QIS) is an emerging interdisciplinary field at the intersection of physics, computer science, electrical engineering, and mathematics leveraging the laws of quantum mechanics to circumvent classical limitations on information processing. With QIS coursework proliferating across US institutions, including at the undergraduate level, we argue that it is imperative that ethics and social responsibility be incorporated into QIS education from the beginning.
We discuss ethical issues of particular relevance to QIS education that educators may wish to incorporate into their curricula. We then report on findings from focus interviews with six faculty who have taught introductory QIS courses, focusing on barriers to and opportunities for incorporation of ethics and social responsibility (ESR) into the QIS classroom. Few faculty had explicitly considered discussion of ethical issues in the classroom prior to the interview, yet instructor attitudes shifted markedly in support of incorporating ESR in the classroom as a result of the interview process itself. Taking into account faculty's perception of obstacles to discussing issues of ESR in coursework, we propose next steps toward making ESR education in the QIS classroom a reality.


\section*{Introduction and background}

Ethics education is increasingly being recognized as an integral element of STEM education. Engineering disciplines have long recognized the importance of ethics in education, with the ASEE issuing a statement on the importance of explicitly addressing ethics education in the classroom as early as 1999 \cite{ASEE:1999}. ABET lists 7 documented student outcomes to be met in baccalaureate engineering programs, including ``an ability to recognize ethical and professional responsibilities in engineering situations and make informed judgments, which must consider the impact of engineering solutions in global, economic, environmental, and societal contexts'' \cite{ABET:2020}. Physics is also beginning to embrace conversations on ethics, partly in response to a 2004 statement in \textit{Physics Today} calling for ethics education in response to data fabrication scandals \cite{Kirby:2004}. In April 2019, the American Physical Society (APS) released an official statement entitled Guidelines on Ethics calling for, among other provisions, integrating ethics education into the training of physics students \cite{APS:Ethics:2019}. Two months later, the APS Ethics Committee held its first meeting \cite{Poffenberger:2019}; its website has since become a hub for STEM ethics education materials. 

\subsection*{A brief overview of QIS education and QIS education research}

Quantum information science (QIS) is an emerging interdisciplinary field at the intersection of physics, computer science, electrical engineering, and mathematics leveraging the laws of quantum mechanics to circumvent classical limitations on information processing. The National Science and Technology Council's Subcommittee on Quantum Information Science identifies four key areas of fundamental research within QIS: quantum sensing, quantum computing, quantum networking, and broader scientific advances enabled by advances in quantum theory and devices \cite{SCQIS:2018}. Fueled in part by the National Quantum Initiative Act of 2018, interdisciplinary QIS courses at the undergraduate or combined undergraduate/graduate level have  begun to proliferate across US universities \cite{Fox:2020}. Recognizing the need to invest in a robust QIS educational program in the US for the ``quantum revolution,'' 32 scientists and professionals in QIS and adjacent fields signed an open letter calling for the creation of additional coursework and degree programs in QIS, including the early involvement of education experts in curriculum development \cite{Aiello:2021}. 

Due to the relative novelty of QIS as a topic of coursework, physics education research (PER) and discipline-based education research (DBER) more broadly have only recently begun to explore the field of QIS education. Early research in physics and engineering education focused on identifying ways to teach quantum mechanics to computer scientists and engineers for quantum information applications \cite{Mermin:2003,Grau:2004}. More recent work has sought to present a portrait of QIS education as it exists today \cite{Plunkett:2020,Seegerer:2021,Cervantes:2021,Meyer:2022} and to study student reasoning in QIS courses \cite{Meyer:2021}. Beyond DBER, specific ethics issues related to quantum technology and its applications have also been discussed extensively in the literature \cite{Coenen:2017, deWolfe:2017, EPSRC:2018, Hibbler:2019, Holter:2021, Inglesant:2021, Perrier:2021}; however, these studies typically examine QIS ethics from the researcher or policymaker's perspective rather than the educator's.

Our work aims to fill the gap between these respective strands of inquiry by addressing faculty attitudes and perceived opportunities and barriers to incorporating discussion of ethics and social responsibility (ESR) in the classroom. We believe that the QIS educator's positionality and perspective are different enough from that of the QIS research community to merit separate study, particularly given that our work suggests many educators teaching university QIS courses are not themselves experts in the field \cite{Meyer:2022}. Our study thus complements recent work studying ``responsible quantum'' discourse among QIS researchers and industry stakeholders \cite{Roberson:2021Arxiv}, whose perspectives likewise strongly drive ethics narratives in the quantum community.

\subsection*{Ethics education in physics}

Discussions of ethics education in physics have traditionally lagged behind those in engineering disciplines.
While engineering education has a strong record of successfully incorporating ethics into the classroom, we focus primarily on ethics education in physics for three reasons. First, as physicists, our own positionality grants us unique access to assess the state of ethics education within our own discipline. Second, though QIS is a notably interdisciplinary field, a preponderance of QIS courses at U.S. institutions appear to be listed in physics departments and/or taught by physics faculty \cite{Cervantes:2021,Meyer:2022}, suggesting that physics cultures have a disproportionate influence on practices of QIS teaching. Finally, QIS education programs do not yet fall under the purvey of ABET accreditation, whose standards \cite{ABET:2020} are undeniably a major impetus for ethics education in many engineering disciplines.

Nevertheless, recent work in PER and the engineering education community has built scaffolding for incorporating ethics into the physics curriculum. 
A 1998 study found that, despite concerns about feasibility given departmental climate, support was strong among physicists in both academia and industry for inclusion of a mandatory or elective ethics courses into undergraduate and graduate physics education \cite{Wylo:1998}.
Eastern Michigan University's physics department has offered a successful course on ethical issues in physics since 1988 \cite{Thomsen:2007}. More recently, PER researchers studied the impact of a 2-week unit on the development of the atomic bomb in a modern physics class on students' development and application of ethical principles, showing that many students apply productive approaches toward the development of ethical arguments and benefit from a strong scaffolding in ethical theory even if they sometimes struggle to correctly apply the theory \cite{Ochoa:2019}. The team also found that practices designed to elicit students' ethical opinions and develop them through engagement with other students can  bolster students' perception of agency in ethical conversations within physics \cite{Gutmann:2020}. Parallel attempts to integrate issues of equity and inclusion into the physics classroom have shown similarly promising results 
\cite{Daane:2017, Dalton:2020, Hoehn:2020, Baylor:2021}. And preliminary evidence suggests that ethics case studies are effective in reducing unethical student behaviors in physics lab courses \cite{Conry:2020}.

\subsection*{QIS courses as sites for ethics education}

As the importance of ethics education becomes more accepted within physics and engineering, we believe that QIS courses are particularly ripe sites to target for interventions in ESR for three reasons. First, a primary obstacle to STEM ethics education has historically been the need to retrofit these topics into established curricula. With the rapid growth in QIS coursework at US institutions \cite{Aiello:2021}, the rare opportunity exists to incorporate ESR as essential components of QIS education from a discipline's beginning. Second, as an interdisciplinary field at the intersection of physics, computer science, electrical engineering, and mathematics, QIS courses provide the opportunity for faculty from diverse disciplinary backgrounds to teach collaboratively. As a result, QIS courses provide an ideal opportunity to facilitate transfer of ethics education techniques -- and acceptance of ethics education in the classroom more broadly -- from disciplines where the importance of ethics education is relatively accepted to disciplines such as physics that have historically lagged behind. Third, QIS courses bring together students of diverse academic backgrounds \cite{Meyer:2022}, a boon for ethics education given the multiplicity of values students are likely to bring to the course.

Incorporating topics of ESR into the QIS classroom will also have downstream impacts on the quantum industry as a whole. Roberson's recent study of discourse surrounding ``responsible quantum''  highlights the importance of shared language regarding social responsibility and ethics in innovation \cite{Roberson:2021Arxiv}. Discussing topics of ESR in the classroom will help students develop the common language  necessary for these conversations to thrive across academia and industry.

\subsection*{Specific ethical issues pertaining to quantum technology}

As with any engineering or applied science field, ethical issues abound in QIS; there is no way we can compile a complete listing of such issues. However, based on coverage in media and the literature and on our own exposure to discourse within the field of QIS, we identified four key ethical issues in QIS education and quantum technology as a starting point for our project. We emphasize that this list is not intended to be a complete inventory of ethical issues in QIS, but simply represents a subset that the research team found particularly obvious and salient:

\begin{itemize}
    \item \textbf{Cybersecurity} -- RSA encryption, which forms the backbone of today's internet security protocols, relies upon the extreme difficulty of factoring large integers. Interest in quantum computing as a field surged in 1994 when P.W.\ Shor published an algorithm for factoring integers in polynomial time \cite{Shor:1994}.
    Suddenly quantum computing emerged as a threat to individual privacy and national security. Cryptographers have since begun to develop practical, quantum-resistant classical encryption schemes to supplant RSA encryption \cite{Buchanan:2017}. Though secure quantum communication channels \cite{Bennett:1984,Bennett:1992,Liao:2017} might revolutionize security for particularly sensitive data such as state secrets, an even greater undertaking will be to retrofit society's vast network of classical communication channels to quantum-proof encryption scheme. How can we prepare society for a post-quantum internet, protecting individuals and businesses from quantum-enabled surveillance and cybercrime?
    \vspace{10pt}
    \item \textbf{Military applications} -- Quantum technologies have significant applications in defense and weapons technology, from position, navigation, and timing in GPS-denied environments to detection of underground structures to possible advances in artificial intelligence \cite{Ingelsant:2018,Krelina:2021,Neumann:2021,Sayler:2021}. Accordingly, both the US and Chinese militaries have invested strongly in fundamental and applied research in quantum technologies. The rapid growth in funding has led to fears of a quantum arms race between world powers \cite{EPSRC:2018,Lele:2021,Roberson:2021Arxiv}. Do QIS practitioners have the responsibility to advocate for quantum technology to be used for peaceful purposes, and to consider the ethical implications of military funding on their research \cite{Papadopoulos:2008}? 
    \vspace{10pt}
    \item \textbf{Rhetoric and media coverage of quantum technologies} -- Our work has identified concern that quantum technology is overhyped in the media as a particularly salient issue for faculty teaching QIS courses.\footnote{We anticipate this finding will be the subject of a future publication.} Media hype of quantum technology may have unwanted consequences for international policy \cite{Smith:2020} and the business community. Concerns have also been raised about unintended consequences of terms such as ``quantum supremacy'' \cite{Palacios:2019} and whether the rhetoric of quantum technology in the media favors international competition over public good \cite{Roberson:2021QST}. The rhetoric of QIS practitioners both shapes and responds to the so-called media ``hype cycle'' \cite{Roberson:2021Minerva,Fenn:2008}. Indeed, with the current media focus on quantum technology, quantum scientists are arguably in an ideal position to shape public discourse \cite{EPSRC:2018}. How can scientists communicate an accurate picture of quantum technology grounded in science rather than speculation? Might there be ways to leverage media hype to promote scientific literacy or build public support for scientific research?
    \vspace{10pt}
    
    \item \textbf{Equity and inclusion in quantum workforce development} -- Multiple studies in the quantum community has focused on the need for workforce education and development to drive the coming ``quantum revolution'' \cite{Fox:2020,Aiello:2021,Hughes:2021}. Without careful attention to equity and inclusion, however, QIS runs the risk of replicating the patterns of its progenitor fields, which have made among the least progress among STEM fields in tackling longstanding diversity problems \cite{APS:race,APS:gender}. To combat these trends, Aiello \textit{et al.} call specifically for an emphasis on programs to promote retention, professional development, and QIS identity formation for women and underrepresented racial and ethnic minorities \cite{Aiello:2021}. As with any technological revolution, there are also questions of how to ensure that the benefits of quantum technology do not just accrue to the largest corporations and wealthiest nations \cite{EPSRC:2018}. How can we ensure that wealth, jobs, and educational opportunities generated by the quantum industry are equitably distributed? And what will happen to these jobs should commercial applications of quantum technology prove unfeasible?
    \vspace{10pt}
\end{itemize}

Of course, QIS necessarily touches on many more general issues of research ethics such as plagiarism, data integrity, and abuse of publication metrics. Though we acknowledge the importance of such issues for QIS research, we limit our focus of this paper to those issues not traditionally treated by more general discussions of STEM ethics.

\begin{table*}[tb]
    \centering
    \resizebox{\textwidth}{!}{%
    \begin{tabular}{c c c c c c c c}
        \hline \hline
         \thead{Course} & \thead{Pseudonym} & \thead{Home Dept} & \thead{Institution Type} & \thead{Course Level} & \thead{Listed Dept(s)} &  \thead{Since}  \\
         \Xhline{2\arrayrulewidth}
         \thead{A} & Albert & Computer Science & Private R1 & BFY Undergrad & Computer Science & 1999\\
         \thead{B} & Ben & Physics & Private R1 & Hybrid & Physics & 2020\\ 
         \thead{C} & Carl & Physics & Public R1 & BFY Undergrad & \makecell{Physics, Computer Science} & 2020\\
         \thead{D} & David & Physics & Public R1 & Hybrid & \makecell{Physics, Computer Science} & 2011\\
         \thead{E} & Edwin & Physics & Private R1 & Hybrid & \makecell{Physics, Computer Science} & 2000*\\
         \thead{F} & Franz & Math & \makecell{Private Baccalaureate} & BFY Undergrad & Computer Science & 2018\\
         \hline \hline
    \end{tabular}}

    \caption{Demographic profiles of the six courses and their instructors for which follow-up interviews were conducted. Institution type is from the Carnegie Classification of Institutions of Higher Learning \cite{carnegie}; all information self-reported by instructors. ``Physics'' includes engineering physics. BFY indicates courses intended for an audience of beyond-first-year undergraduates.
    *Date approximate because course predates current instructor, who began offering it in 2008.}
    \label{tab:interviewees}
\end{table*}

\section*{Methodology}

This work is situated as the first step in a broader project to incorporate ESR into the QIS classroom. Our aim for this preliminary work is therefore twofold: (1) to assess the value and feasibility of efforts to incorporate ethics education into the QIS classroom, and (2) to identify key opportunities and obstacles to incorporating ethics education into the QIS classroom. The second objective is motivated by research in the engineering ethics community showing the importance of faculty and institutional engagement for successful ethics education in the classroom 
\cite{Walczak:2010, Polmear:2018, Tang:2018, Fatehiboroujeni:2019}. 

We conducted interviews in summer 2021 with 6 faculty members who had taught introduction to QIS courses at the undergraduate or hybrid undergraduate/graduate level at a US institution. Interviews were conducted in person or over Zoom depending on geographic proximity. Though the interview protocol focused primarily on technical education in QIS, our analysis in this paper specifically focuses on instructors' responses to the final segment of the interview, which focused on faculty experiences incorporating ESR into the QIS curriculum. This interview segment was placed last so as to minimize undue influence on the remainder of the interview questions.

The backgrounds of the 6 faculty interviewees and the interviewee selection process are profiled in detail in \cite{Meyer:2022}. Table~\ref{tab:interviewees} shows more detailed information on the QIS course each interviewee teaches. Interviewees were selected based on their responses to a survey sent out to QIS instructors in spring 2021; the majority of survey recipients were identified through an analysis of course catalogs as outlined in Ref.~\cite{Cervantes:2021} with the remainder recruited at a conference for undergraduate QIS education.

After a brief overview of what we meant by ``ethical and social issues'' and a few examples, we asked faculty to elaborate on the following semi-structured interview prompts:

\begin{itemize}
\item \textit{Do you engage in any discussion of these issues (formal or informal) with your students? Why or why not? (If so, how?)}
\item \textit{Do you believe it is appropriate (and valuable) to engage in a discussion of these issues with QIS students in a course context? Why or why not?}
\item \textit{If you would be interested in incorporating QIS ethics discussions into your course but do not do so already, what if any are the obstacles? Are there resources we could provide that would be helpful?}
\end{itemize}

Barriers noted by faculty were classified using a hybrid coding scheme with two layers. First, we analyzed instructor responses to identify specific barriers, opportunities, and attitudes relating to discussion of ESR in the QIS classroom. Second, we classified barriers as personal or structural in relation to the instructor's experience. We define personal barriers as relating to the instructor's own attitudes, beliefs, and skills -- in other words, factors that are largely under the control of the instructor or instructional team. We define structural barriers as those fundamentally rooted in the incentives structure of the university or academic discipline. We also qualitatively examined faculty attitudes for shifts over the course of the interview process using discourse analysis \cite{Edwards:1992} of their interview responses.

\section*{Results}

\subsection*{Barriers to discussion of ethics and social issues in QIS courses}

Consistent with prior work in the engineering ethics literature (e.g.\ \cite{Polmear:2018}), we find that faculty perceive significant barriers to the incorporation of ethics and social issues in the QIS classroom. For the purposes of discussion, we classify these barriers as personal or structural as defined above. We acknowledge this distinction is far from entirely black-and-white, but we find it useful for classifying and interpreting these obstacles in terms of the primary sites for intervention needed to overcome them.

Personal barriers stem primarily from faculty's views and perceptions of their own role and skills as instructors. Specific personal barriers identified by faculty are discussed below:

\begin{itemize}
    \item \textbf{Role as QIS instructors:} Some faculty felt, at least early on in the discussion, that teaching ethical thinking was beyond the scope of their role as technical instructors. A common refrain was that ESR may be important but is best sited in non-technical disciplines:
    
    \begin{iquote}
    \who{David}
    What the motivations of various people in society are for developing this technology, what the use is it could be put to ... I guess I never really felt it was my role to read the script.
    \timestamp{1:25:00 verified}
    \end{iquote}
    
    Carl likewise argued that though ethical issues undeniably emerge in quantum technology, they do not diverge significantly from those of other areas of engineering and technology and are perhaps better suited for philosophy departments, an attitude also expressed by multiple participants in the interview study of QIS researchers and stakeholders discussed in the introduction \cite{Roberson:2021Arxiv}. 
    \vspace{10pt}
    \item \textbf{Perceived lack of personal expertise: } A common refrain was that faculty felt they lacked the expertise needed to teach ESR effectively in the classroom. Faculty perhaps viewed their own technical education as leaving them ill-prepared for discussing these issues:
    
    \begin{iquote}
    \who{Edwin}
    I don't regard this [ESR] as anything that I have much to add.\timestamp{1:19:45 verified}
    \end{iquote}
    \vspace{-0.4\baselineskip}
    \begin{iquote}
    \who{Ben}
    I wouldn't be an expert on it [ESR], so I kind of leave that to the experts. You know, there's ... an argument for, well, we need to become experts if we're teaching this.
    \timestamp{47:00 verified}
    \end{iquote}

    Ben further relates how he was considering teaching a course on the lives of African American physicists but quickly ran into the obstacle of lack of subject-matter expertise:
    
    \begin{iquote}
    \who{Ben}
    We would read books ... and there's supposed to be an anti-racist subtext. But critical race theory is a real academic discipline that I know nothing about. So I wouldn't attempt to discuss it explicitly ... Back to quantum information science, if there was a way to teach the physics with some kind of social justice subtext, that would be great, but I'd -- I don't really know how to engage with that material head-on.
    \timestamp{48:40 verified}
    \end{iquote}
    
    In addition, many QIS instructors do not even identify as experts in technical aspects of QIS \cite{Meyer:2022}, much less ethical ones. Franz suggests that some of his discomfort discussing ESR in the classroom more generally stems from his positionality as a mathematician teaching to a computer science audience:
    
    \begin{iquote}
    \who{Franz}
    In general, in my computer science courses, I think I'm, I'm weak on ethical issues ... because I'm an outsider, because I'm a mathematician, and I think I'm better about talking about them in math topics.\timestamp{1:20:45 verified}
    \end{iquote}
    \vspace{-.4\baselineskip}

    \vspace{10pt}
    
    \item \textbf{Perceived lack of facilitation skills:} In addition, Ben expressed wariness toward leading discussions of ESR in the classroom because of given perceived lack of experience facilitating conversations on controversial issues -- a skill that STEM faculty in particular may never have had the chance to develop:
    
    \begin{iquote}
    \who{Ben}
    If some student, for example, says something offensive to other students ... I have zero skill or experience in facilitating discussions -- I would be helpless. \timestamp{51:45 verified}
    \end{iquote}

    \vspace{-.4\baselineskip}

\end{itemize}
\vspace{10pt}

The personal barriers identified above can be addressed through increased exposure to ethical issues and proper facilitation training. However, another class of barriers has to do with factors outside faculty's immediate control and may require systemic cultural change to overcome. Interviewees identified the following structural barriers to ESR in the QIS classroom:

\begin{itemize}
    \item \textbf{Limited class time:} A common concern was that incorporating issues of ESR into the QIS classroom would further reduce the already limited amount of time available in the course for core technical content. As many as 5 interviewees expressed consternation that their typically 1-semester elective courses already didn't include enough time to cover the technical content they wished. As such, taking time out of the class (even for a few lectures) to talk about ethical issues was a difficult sell for some faculty:
    
    \begin{iquote}
    \who{David}
    I wouldn't want to spend too much time [on ESR] because, again, time's limited in the class.\timestamp{1:25:50 verified}
    \end{iquote}
    \vspace{-0.4\baselineskip}
    \begin{iquote}
    \who{Edwin}
    I have a finite amount of time in the course. There's only so much you can cover.\timestamp{1:19:37 verified}
    \end{iquote}
    \vspace{-.4\baselineskip}
    \vspace{10pt}
    
    \item \textbf{Politicization and the need to take a neutral perspective:} Edwin expressed concern about bringing contentious issues into the classroom given the political polarization of today's college campuses. Edwin explains that he has been shying away from such discussions in the classroom precisely because of the current political climate:
    
    \begin{iquote}
    \who{Edwin}
    It's especially problematic now given ... the polarization we've gone through ... I tried to avoid, for the most part, getting into any of these issues, you know, during the [recent] era.\timestamp{1:23:40 verified}
    \end{iquote}
    
    Though political polarization does not necessarily preclude discussions of ESR in the classroom, such an environment does require such discussions to be approached tactfully. David expressed confidence that it was possible to discuss issues of ESR in the QIS classroom without provoking partisan hostility given his own experience in other courses. Discussing politically charged issues, in his view, does not necessarily require adopting a `neutral' stance -- a perhaps impossible standard -- but rather an open-minded one that encourages all students to develop critical thinking skills and self-efficacy. In his words:
    
    \begin{iquote}
    \who{David}
    I always feel constrained to -- not so much to be neutral, but to be studiously open to different points of view, ... teach tools, in this case like a mode of thinking ... -- or maybe it's not even a mode, just ... instilling in the students the confidence that they can think about these things on their own. \timestamp{1:30:59 verified}
    \end{iquote}
    
    David later explains that to him, being open means ``not so much neutral as accepting, listening -- so to speak -- and curious.'' David's experience shows that politicization in the classroom can be managed if met with openness and intellectual humility.
    
\end{itemize}
\vspace{10pt}

Interestingly, none of the faculty expressly cited potential pushback from their departments as a barrier to incorporating issues of ESR into the QIS classroom. Perhaps QIS instructors face less pressure to adhere to narrow definitions of their academic discipline than physicists given the explicitly interdisciplinary audience many instructors are teaching for \cite{Meyer:2022}. Alternatively -- as discussed below -- many faculty had simply never explicitly considered incorporating ESR into their QIS curricula as discussed below, so another explanation may be that faculty simply have no basis to gauge possible reactions from colleagues who, in all likelihood, are equally unfamiliar with the idea.

\subsection*{Conversation alone broadened faculty attitudes toward ESR education}

Perhaps the greatest takeaway from our study was the extent to which faculty's attitudes on ESR in the classroom evolved throughout the course of our discussion. Even though this section of the interview typically lasted only 5-15 minutes at the end of a much longer interview, certain faculty's views on the appropriateness and feasibility of incorporating issues of ESR into the classroom shifted markedly toward openness to ESR in the classroom during the course of our discussion -- a shift most pronounced for Carl and David, whose cases we discuss below. We find that mere exposure to the idea of incorporating ESR into the QIS classroom, along with empathetic listening and questions encouraging deeper reflection, often was sufficient to prompt a noticeable change in faculty's attitudes regarding ESR in the QIS classroom.

One reason our conversations might have had this surprising effect was simply that multiple faculty had never thought explicitly about teaching ESR in the classroom (at least in those terms). David's change in attitude seemed to have been sparked simply by the interview topic itself:

\begin{tightquote}
\who{David}
I always thought of this quantum computing course as a very technical-type course. Never thought of it in those terms [ESR] but ... it could be an opportunity. \timestamp{1:28:09 verified}
\end{tightquote}

Further, we argue that the way we presented these ideas in the interview -- simply asking faculty to reflect and elaborate on their views from a non-judgmental position -- itself was a powerful agent of change. Carl began the interview reasonably skeptical of ESR in the classroom. However, a follow-up question intended simply to clarify one of his responses prompted a shift in his perspective from one of skepticism to general agreement:

\begin{tightquote}
\who{Carl}
I mean, there's certain problems that come up about quantum computers. But I'm sure they are examples [of] much more general questions that are best -- best addressed by the philosophy department.
\who{Interviewer} Yeah, I guess I would follow up on that though ... just because they're discussed in philosophy departments doesn't mean that students are taking the philosophy classes, they're taking your class ... Do you think that there are any benefits to having it in the same course ... having it be a part of quantum computing education? 
\who{Carl}
I have never thought of it! Let me just, can I? -- I'll just think for a minute. Yeah, I think there could be benefit if there was a sort of clear discussion of ... what the proposed technology is for and what good and bad uses it could be put to. So I guess that's more domain specific. \timestamp{1:30:20 verified}
\end{tightquote}

In the course of this dialogue -- thanks to the follow-up question posed by the interviewer -- Carl seems to have evolved in his thinking. Even though the interviewer had previously mentioned a handful of ethical and social issues in QIS in order to start the conversation, it wasn't until actually wrestling with the interviewer's follow-up question above that Carl realized there could be value in discussing discipline-specific ethical issues in the QIS classroom. 

The finding that the interview process itself shaped instructor attitudes toward incorporating ESR into the QIS classroom was somewhat surprising: we expressly designed our interview protocol simply to document interviewees' views without unduly influencing them. Throughout the course of the interviews, we sought to keep an open-minded, nonjudgmental perspective, and we reminded faculty that there were no right or wrong answers to the questions we asked. Yet Carl and David displayed marked changes in the tone of their responses during the course of this section of the interview in spite of our precautions. Clearly, the interview process itself, despite attempting to take no stance on the research questions being asked, was sufficient to increase faculty receptivity toward the idea of incorporating ESR into the QIS classroom. Rather than interpreting these faculty interviews as simply vehicles to elicit interviewees' perspectives, we must acknowledge that interviewees' perspectives and mental models are apt to shift over the course of the interview even if the interview questions are written to be neutral. As prior work in PER has shown, the interview is itself a dynamic process involving interaction with the interviewer that can form a ripe site for broader reconceptualization of one's own understandings \cite{Rebello:2004, Sherin:2007, Corpuz:2011}.


\section*{Discussion and conclusions}

The preliminary work presented in this paper is, by itself, necessary but insufficient to transform the cultures of QIS education. To have an impact on students, the lessons from this study must translate to actual interventions bringing ethics to the QIS classroom. Given faculty members' lack of experience discussing ethical issues in the classroom as reported here and in the literature \cite{Walczak:2010,Polmear:2018} and the limited time available for ethics education in QIS courses, we echo Ref.~\cite{Dyrud:2015} in recommending the development of modular resources centered on ethics case-studies, coupled with guidance for leading discussions and tie-ins with ethical theory, as in Ref.~\cite{Ochoa:2019}. Ben and David indicated they would like to see resources such as curated readings:

\begin{tightquote}
\who{Ben}
If there was some guidance -- like, this is what you should learn about the associated ethical and social issues -- then I'd look into it. \timestamp{47:20 verified}
\end{tightquote}

\begin{tightquote}
\who{David}
If you produce something like a set of curated readings on this ... that by itself would probably be a very useful resource. \timestamp{1:28:55 verified}
\end{tightquote}

We believe that the barriers mentioned by our interviewees in the previous section can be resolved, if not for all faculty then at least for those who are willing to approach the concept with an open mind. Ben and David's idea of a set of curated readings and resources is a practical starting point: Curated resources could help faculty overcome a perceived lack of personal experience, and associated facilitation guidelines could assist faculty who feel they lack the necessary facilitation skills or are concerned about political polarization in the classroom. Such resources could also alleviate instructor concerns that discussing ESR would consume too much class time, particularly if these resources are presented in a modular format adaptable to the time available, whether that be 15 minutes or several entire lecture periods. Finally, simply providing and advertising such a resource list could help more faculty become aware of quantum-specific ethical issues and the concept of incorporating ESR into the QIS classroom.

We argued above that QIS courses are an ideal site for incorporating ESR into the physics and engineering classroom. We also believe strongly that with the appropriate attitude and motivation, any faculty member has the tools to engage in valuable discussions of ESR in their QIS course. We thus believe there is an urgent need for education experts to devise resources for faculty interested in incorporating these issues into the QIS classroom, and to prioritize doing so in the early years of the ``quantum revolution'' while paradigms and philosophies in QIS education remain in flux. At the same time, we encourage educators to try incorporating ethical and social issues into their QIS curricula, even if it is as simple as bringing in a contrary perspective on the technological value of quantum key distribution (as Franz does) or discussing the implications of Shor's algorithm for internet security (like David). Incorporating ESR into the QIS classroom doesn't necessarily mean revamping curricula -- it merely requires creating the time and space to discuss and critically examine these issues rather than treating them as tangential asides.
If we instead brush social issues under the rug in our courses, we risk alienating the very students we wish to recruit to become the next generation of QIS professionals. In Franz's words:

\begin{tightquote}
\who{Franz}
It's a luxury, it's a privilege to be able to think about abstract problems that are divorced from society ... you can only do that if you're in a state of extreme privilege ... [it's] counterproductive for us to be ignoring most of the population who could be contributing!\timestamp{1:26:20 verified}
\end{tightquote}

\subsection*{Limitations and next steps}

Though our work here is among the first to examine integration of ESR into the QIS classroom, we acknowledge our work is but a start. We hope this preliminary study will be the first step toward incorporating ESR into QIS classrooms at our institution and beyond. In the meantime, we identify two key limitations of this study that ought to be addressed in future work.

First, though QIS faculty perspectives are undoubtedly important in developing effective resources for ethics education, they are far from the only voices that matter. All of the QIS faculty interviewed here have relatively limited experience, if any, discussing ethical issues in the QIS classroom. The involvement of faculty, particularly in engineering, who have extensive experience incorporating ethics into STEM classrooms will be particularly necessary to ensure lessons from ethics education in adjacent disciplines are transferred to QIS education. Likewise, student perspectives on ethics education will be necessary to ensure that curricular materials developed are truly relevant to students' lives.

Second, the perspectives highlighted in this work are somewhat unrepresentative of QIS educators as a whole. Our interviewees identified exclusively identified as white and male and were disproportionately housed in physics departments at R1 institutions -- faculty views on ethics education are shown to be influenced by gender and departmental affiliation \cite{Katz:2017}. And response bias likely skews the interview pool toward those open to research-based teaching methods.

From the start, ESR education in QIS ought to incorporate and respond to recent criticism of traditional STEM ethics education practices. While engineering ethics education has traditionally focused on helping students develop ethical reasoning skills, a strong argument can be made that the ultimate goal of ethics education ought to be the cultivation of more ethical \textit{behaviors} and should be designed and assessed accordingly \cite{Clancy:2021}. Another criticism is that engineering ethics education tends to focus on microethics (the actions of the individual researcher or engineer) over macroethics (pertaining to matters of society) \cite{Dyrud:2015} when many of the most important ethical issues in QIS today (e.g.\ cybersecurity) far transcend the actions of any individual researcher.

Finally, it is valuable to recognize that much of classical ethical theory is rooted in Eurocentric theories of moral philosophy that -- though undoubtedly important -- may not be equally applicable to students of all backgrounds. Indeed, culturally-specific perspectives on ethical issues have been identified as potential barrier to the retention of Native American and Alaska Native students in engineering and STEM \cite{Ingram:2021}. In order to engage students of diverse identities and backgrounds, ESR education in QIS ought to include space for counternarratives. Broadening definitions of research ethics to incorporate critical theories and social justice narratives may be one way to help bridge this gap and ensure that ethics lessons in QIS are culturally appropriate for the diversity of researchers and engineers we seek to train \cite{Jalali:2019}.

\section*{Acknowledgments}

Special thanks to Bianca Cervantes, Gina Passante, and Steven Pollock for their work leading up to this interview study, and to our faculty interviewees. This research is funded by the CU Boulder Department of Physics, the NSF GRFP, and NSF Grants No.'s 2012147 and 2011958.

\vspace{4\baselineskip}\vspace{-\parskip} 
\footnotesize 
\bibliographystyle{ieeetr}


\end{document}